\documentclass[11pt]{article}

\usepackage{hyperref}

\usepackage{amssymb}
\usepackage{amsmath}
\usepackage{bm}
\usepackage{latexsym}
\usepackage{url}
\usepackage{cite}

\newtheorem{theorem}{Theorem}

\newtheorem{cor}[theorem]{Corollary}

\newcommand{\microspace}{\mspace{0.5mu}}

\newcommand{\tr}{\operatorname{Tr}}
\newcommand{\pt}{\operatorname{T}}

\newcommand{\ip}[2]{\left\langle #1 , #2\right\rangle}
\def\({\left(}
\def\){\right)}
\def\Id{\bm{1}}

\newcommand{\setft}[1]{\mathrm{#1}}
\newcommand{\lin}[1]{\setft{L}\left(#1\right)}

\newcommand{\herm}[1]{\setft{Herm}\left(#1\right)}

\def\complex{\mathbb{C}}

\def \lket {\left|}
\def \rket {\right\rangle}
\def \lbra {\left\langle}
\def \rbra {\right|}
\newcommand{\ket}[1]{\lket\microspace #1 \microspace\rket}
\newcommand{\bra}[1]{\lbra\microspace #1 \microspace\rbra}

\def\A{\mathcal{A}}
\def\B{\mathcal{B}}

\begin{document}

\title{Small sets of locally indistinguishable orthogonal maximally entangled states}

\author{
  Alessandro Cosentino\hspace{5mm}
  Vincent Russo\\[2mm]
  \emph{\small School of Computer Science and Institute for Quantum
    Computing}\\
  \emph{\small University of Waterloo, Canada}
}

\date{April 10, 2014}

\maketitle

\begin{abstract}
We study the problem of distinguishing quantum states 
using local operations and classical
communication (LOCC).
A question of fundamental interest is whether
there exist sets of $k \leq d$
orthogonal maximally entangled states in $\complex^{d}\otimes\complex^{d}$
that are not perfectly distinguishable by LOCC.
A recent result by Yu, Duan, and Ying [Phys. Rev. Lett. 109 020506 (2012)]
gives an affirmative answer for the case $k = d$.
We give, for the first time, a proof that such sets of states
indeed exist even in the case $k < d$.
Our result is constructive and holds for an even wider class of operations 
known as positive-partial-transpose measurements (PPT).
The proof uses the characterization of the PPT-distinguishability problem
as a semidefinite program.
\end{abstract}

\section{Introduction}
A central subject of study in quantum information theory is 
the interplay between entanglement and nonlocality.
An important tool to study this relationship is the paradigm
of local quantum operations and classical communication (LOCC). 
This is a subset of all global quantum operations, 
with a fairly intuitive physical description. 
In a two-party LOCC protocol, 
Alice and Bob can perform quantum operations only on their local
subsystems and the communication must be classical.
This restricted paradigm has played a crucial role in the understanding 
of the role of entanglement in quantum information. It has also provided a framework 
for the description of basic quantum tasks such as quantum key distribution 
and entanglement distillation.

A fundamental problem that has been studied to understand the limitations of 
LOCC protocols is the problem of distinguishing quantum states.
The setup of the problem is pretty simple in the bipartite case.
The two parties are given a single copy of a quantum state chosen
with some probability from a collection of states and their goal is to identify which state was given, 
with the assumption that they have full knowledge of the collection.
If the states are orthogonal and global operations 
are permitted, then it is always possible to determine the state
with certainty. In contrast, if only LOCC protocols 
are allowed, Alice and Bob cannot in general discover the state they 
have been given, even if the states are orthogonal.
The problem of distinguishing among a known set 
of orthogonal quantum states by LOCC has been studied by several researchers 
\cite{Bennett99,Walgate00,Ghosh2001,Walgate2002,
Horodecki2003,Fan04,Ghosh04,Nathanson05,Watrous2005,
Owari2006,Yu2011,Yu2012,Yu2012a,Cosentino2013,Bandyopadhyay2013}.
Some direct applications of this problem include secret sharing 
\cite{Gottesman2000} and data hiding \cite{DiVincenzo2002}.

A question of basic interest is how the size of LOCC-indistinguishable sets 
(denoted by $k$ in this paper) relates to the local dimension $d$ of each of 
Alice's and Bob's subsystems.
We know that the dimension of a quantum system puts a bound on the degree 
of entanglement the system could possibly have with another system.
Analogously, one can ask whether the local dimension of the two subsystems 
plays any special role in the nonlocality exhibited by  
LOCC-indistinguishable sets of states.

Walgate et al. \cite{Walgate00} proved that any two orthogonal 
pure states can always be perfectly distinguished by 
an LOCC measurement.
A particularly interesting case is when the set is constituted of orthogonal states 
with full local rank.
Regarding this case, Nathanson \cite{Nathanson05} showed that it is always possible
to perfectly distinguish any three orthogonal maximally entangled states in 
$\complex^{3}\otimes\complex^{3}$ by means of LOCC.
On the other hand, it is known that $k > d$ orthogonal maximally entangled states 
can never be distinguished with certainty by LOCC measurements \cite{Ghosh04}.
An interesting question is whether there exist sets of $k \leq d$ orthogonal 
maximally entangled states in $\complex^{d}\otimes\complex^{d}$ that are not
perfectly distinguishable by LOCC, when $d > 3$.
For the weaker model of \emph{one-way} LOCC protocols, 
Bandyopadhyay et al. \cite{Ghosh11} showed some explicit examples of indistinguishable 
sets of states with the size of the sets being equal to the dimension 
of the subsystems, i.e., $k=d$.
Recently, Yu et al. \cite{Yu2012} gave an affirmative answer to the question 
for the case $k = d = 4$ in the setting of general LOCC protocols.  
Their result was later generalized in \cite{Cosentino2013} 
for the case $k = d = 2^t$, where $t \geq 2$.
The answer has remained elusive for the case $k < d$.

In this paper, we settle the question by exhibiting, for the first time, 
sets that contain fewer than $d$ orthogonal maximally entangled states in 
$\complex^{d}\otimes\complex^{d}$, which are not perfectly distinguishable by LOCC measurements.
Thus we show that the local dimension of the subsystems is not a tight
bound on the size of sets of locally indistinguishable orthogonal maximally entangled states.

Even though all these results are about maximally entangled states, 
it should be noted that entanglement is not
a necessary feature of locally indistinguishable sets of states. 
In a famous result, Bennett et al. \cite{Bennett99} exhibited 
a set of orthogonal bipartite pure product states that are 
perfectly distinguishable by separable operations, but not by LOCC 
(see \cite{Childs2012} for a simplified proof and a generalization of this result). 
In fact, if we allow states that are not maximally 
entangled to be in the set,
we can construct indistinguishable sets with a fixed
size in any dimension we like.
Indeed, whenever we find a set of indistinguishable maximally entangled states for certain
local dimensions, those states remain indistinguishable when embedded in any larger local
dimensions. Nonetheless they are no longer maximally entangled with respect to the new larger
local dimensions.
On the one hand, entanglement makes distinguishability harder, 
but on the other hand, it can be used as a resource 
by the parties involved in the protocol.
This makes the distinguishability problem especially interesting 
in the case when the set contains only maximally entangled states.

We tackle the problem by studying distinguishability of states for 
a class of operations broader than the class of LOCC measurements,
which is the class of positive-partial-transpose (PPT) measurements.
In fact, this class is even broader than the class of separable measurements, 
for which distinguishability of states has been studied as well 
\cite{Duan2009,Bandyopadhyay2013}.
As opposed to the set of LOCC measurements, 
the set of PPT measurements has a nice mathematical structure. 
Moreover, optimizing over this set is a computationally easy task, 
whereas optimizing over the set of separable measurements is known 
to be an NP-hard problem \cite{Gurvits2003,Gharibian2008}.
Several properties of PPT operations can indeed be characterized in the framework
of semidefinite programming (see \cite{Rains00} for an example).
In fact, semidefinite duality also helps to prove analytical bounds 
on the power of PPT operations, and therefore on the power of LOCC operations.
A straightforward application of this idea is a simplified proof of the 
previously mentioned fact that $k > d$ orthogonal maximally entangled states
cannot be perfectly distinguished by LOCC \cite{Ghosh04}
(see \cite{Yu2012} and \cite{Cosentino2013} for a proof 
that this fact holds for PPT as well).
The characterization of the PPT-distinguishability problem as a semidefinite program 
has been also exploited in \cite{Cosentino2013} to find 
indistinguishable sets with size $k = d$.

A recent work by Yu et al. \cite{Yu2012a} has investigated further
properties of state distinguishability by PPT. They prove a
tight bound on the entanglement necessary to distinguish between three
Bell states via PPT measurements. Furthermore, they show that 
regardless of the number of copies, a maximally entangled state 
cannot be distinguished from its orthogonal complement.

Before giving the definition of a PPT measurement, we review some notation.
We denote by $\A$ and $\B$ the complex Euclidean spaces corresponding to 
Alice's and Bob's systems, respectively.
We assume that $\A$ and $\B$ are isomorphic copies of $\complex^{d}$.
A pure state $u \in \A\otimes\B$ is called maximally entangled 
if $\tr_{\A}(uu^{*}) =\tr_{\B}(uu^{*}) = \Id/d$.
The partial transpose is a mapping on $\A\otimes\B$ defined by 
tensoring the transpose mapping acting on $\A$ and the identity 
mapping acting on $\B$ and it is denoted as $\pt_{\A} = \pt \otimes \Id_{\lin{\B}}$.
Given a complex Euclidean space $\A$, 
we use the symbol $\herm{\A}$ to denote the set
of Hermitian operators acting on $\A$. 
Let $\A = \B = \complex^{2}$ and let $\psi_{i}$, for $i\in \{0, 1, 2, 3\}$, 
be the density operators corresponding to the standard Bell basis, that is,
$\psi_{i} = \ket{\psi_{i}}\bra{\psi_{i}}$, for $i\in \{0, 1, 2, 3\}$, where
\begin{equation}
\label{eq:bell-states}
\ket{\psi_{0}} = \frac{\ket{00}+\ket{11}}{\sqrt{2}},\quad
\ket{\psi_{1}} = \frac{\ket{01}+\ket{10}}{\sqrt{2}},\quad
\ket{\psi_{2}} = \frac{\ket{01}-\ket{10}}{\sqrt{2}},\quad
\ket{\psi_{3}} = \frac{\ket{00}-\ket{11}}{\sqrt{2}}.
\end{equation}

Our construction is based on states that are tensor products of Bell states.
We write down explicitly the action of the partial transpose on the Bell basis:
\begin{equation}
\label{eq:transposebells}
		\pt_{\A}(\psi_{0}) = \frac{1}{2}\Id - \psi_{2}, \quad
		\pt_{\A}(\psi_{1}) = \frac{1}{2}\Id - \psi_{3}, \quad
		\pt_{\A}(\psi_{2}) = \frac{1}{2}\Id - \psi_{0}, \quad
		\pt_{\A}(\psi_{3}) = \frac{1}{2}\Id - \psi_{1}.
\end{equation}
A positive operator $P \geq 0$ is called a 
\emph{PPT operator} if it remains positive under the action of partial transposition, 
that is, $\pt_{\A}(P) \geq 0$. 
A measurement $\{ P_a \geq 0: a \in \Gamma \}$ is called a 
\emph{PPT measurement} if each measurement operator is PPT.

\bigskip
The maximum probability of distinguishing a set of states 
$\{ \rho_{1}, \ldots, \rho_{k}\}$ by PPT measurements 
can be expressed as the optimal value of the following semidefinite program 
(for more details, see \cite{Cosentino2013}).
We are interested in perfect distinguishability,  
so we will assume, without loss of generality, 
that the states are drawn from the set with uniform probability, that is, 
$p_{j} = 1/k$, for each $j = 1, \ldots, k$.

\begin{center}
    \centerline{\underline{Primal problem}}\vspace{-4mm}
    \begin{equation}
    \label{sdp-primal}
    \begin{aligned}
      \text{maximize:}\quad & \frac{1}{k} \sum_{j = 1}^k \ip{P_j}{\rho_{j}}\\
      \text{subject to:}\quad & P_1+ \cdots + P_k = \Id_{\A} \otimes \Id_{\B},\\
      & P_1,\ldots,P_k \geq 0,\\
      & \pt_{\A}(P_{1}), \ldots, \pt_{\A}(P_{k}) \geq 0.
    \end{aligned}
    \end{equation}
\end{center}

The dual of the problem is easily obtained by routine calculation. 

\begin{center}
    \centerline{\underline{Dual problem}}\vspace{-4mm}
    \begin{equation}
    \label{sdp-dual}
    \begin{aligned}
      \text{minimize:}\quad & \frac{1}{k}\tr(Y)\\
      \text{subject to:}\quad & Y - \rho_{j} \geq \pt_{\A}(Q_{j}), \quad j=1,\ldots,k \; ,\\
      & Y\in\herm{\A\otimes\B},\\
      & Q_{1}, \ldots, Q_{k} \geq 0.
    \end{aligned}
    \end{equation}
\end{center}

Given a set of states, an upper bound on the probability of distinguishing 
them by PPT measurements can be obtained by exhibiting a feasible solution
of the above dual problem. 

\section{Main Result}

For any $d \geq 4$ that is a power of $2$, 
we show how to construct sets of $d$ 
orthogonal maximally entangled states in $\complex^{d}\otimes\complex^{d}$, 
for which the above dual problem has optimal value less than or 
equal to $C$, 
where $C < 1$ is a constant.
Given one of such sets, if we consider any of its subsets that contains only $k$ states,
then we have a set of $k$ PPT-indistinguishable maximally entangled states in 
$\complex^{d}\otimes\complex^{d}$, where $k < d$, as long as 
$C < k/d$.
Since any LOCC measurement is a PPT measurement, 
then such a set is also indistinguishable by LOCC. 

\begin{theorem}
\label{th:maintheorem}
For any $d = 2^{t}$, where $t \geq 2$, 
it is possible to construct a set of $k$ maximally 
entangled states in $\complex^{d}\otimes\complex^{d}$ 
for which there exists a feasible solution of 
the dual problem \eqref{sdp-dual}
with value of the objective function equal to $(7d)/(8k)$.
\end{theorem}
\vspace*{12pt}
\noindent
{\bf Proof:}
For the case $t = 2$ ($d = 4$), a set of states was shown 
by Yu et al. in \cite{Yu2012}:
\begin{equation}
	\label{eq:example4}
	\begin{aligned}
		\rho_{1}^{(2)} &= \psi_{0}\otimes\psi_{0},\quad 
    &\rho_{2}^{(2)} &= \psi_{1}\otimes\psi_{1},\quad
		&\rho_{3}^{(2)} &= \psi_{2}\otimes\psi_{1}, \quad
		&\rho_{4}^{(2)} &= \psi_{3}\otimes\psi_{1}.
	\end{aligned}
\end{equation}
This being the first instance in the paper where we use Bell-diagonal states,
we point out that the tensor product structure of those states
should not mislead the reader when considering the cut between 
Alice's and Bob's systems.
If we denote the local systems by $\A = \A_{1}\otimes\A_{2}$ and
$\B = \B_{1}\otimes\B_{2}$, then
the cut is such that the states $\rho_{i}^{(2)}$ lie on the space 
$(\A_{1}\otimes\B_{1})\otimes(\A_{2}\otimes\B_{2})$.

A bound of $7/8$ on the optimal probability of distinguishing these 
states was proved in \cite{Cosentino2013}. 
Here we write the feasible solution of the dual that achieves the value $7/8$:
\begin{align*}
Y^{(2)} &= \frac{1}{4} \Id\otimes\Id - 
	\frac{1}{2}\pt_{\A}(\psi_2\otimes\psi_3), \\
Q_{1}^{(2)} &= \frac{1}{2}[(\psi_0 + \psi_1 + \psi_3)\otimes\psi_2 
	+ \psi_2\otimes(\psi_0 + \psi_1)], \\
Q_{2}^{(2)} &= \frac{1}{2}[(\psi_0 + \psi_1)\otimes\psi_3 
	+ \psi_3\otimes(\psi_0 + \psi_1 + \psi_2)], \\
Q_{3}^{(2)} &= \frac{1}{2}[(\psi_1 + \psi_3)\otimes\psi_3 
	+ \psi_0\otimes(\psi_0 + \psi_1 + \psi_2)], \\
Q_{4}^{(2)} &= \frac{1}{2}[(\psi_0 + \psi_3)\otimes\psi_3 
	+ \psi_1\otimes(\psi_0 + \psi_1 + \psi_2)].	
\end{align*}
By using the set of equations \eqref{eq:transposebells}, 
it is easy to check that the constraints of the dual problem
hold for the above solution.
In fact, it is a straightforward calculation to check that,
for all $j \in \{1, 2, 3, 4\}$, the following equations hold:
\begin{equation}
Y^{(2)} - \rho_{j}^{(2)} = \pt_{\A}(Q_{j}^{(2)}).
\end{equation}
Furthermore, we observe that $Q_{1}^{(2)}, Q_{2}^{(2)}, Q_{3}^{(2)}$, and $Q_{4}^{(2)}$ 
are positive semidefinite, and $\tr(Y^{(2)}) = 7/2$.

For $t \geq 3$, we give a recursive construction of
the states $\rho_{j}^{(t)}$, i.e.,
\begin{equation}
\label{eq:construction}
\rho_{j}^{(t)} =
	\begin{cases}
		\psi_{0}\otimes\rho_{j}^{(t-1)} &\mbox{if } j \leq 2^{t-1},\\
		\psi_{1}\otimes\rho_{j-2^{t-1}}^{(t-1)} &\mbox{if } j > 2^{t-1},\\ 
	\end{cases}
\end{equation}
for $j \in \{1, \ldots, d\}$.
Given this set of states, we can construct, again recursively,
a feasible solution of the dual problem, which achieves the desired bound:
\begin{equation}
\label{eq:generalsolution}
\begin{aligned}
Y^{(t)} 	  &= (\psi_0 + \psi_1)^{\otimes (t-2)} \otimes Y^{(2)}, \\
Q_{j}^{(t)} &= (\psi_0 + \psi_1)^{\otimes (t-2)} \otimes Q_{r}^{(2)}, \quad j \in \{1, \ldots, d\},
\end{aligned}
\end{equation}
where $r \in \{1,2,3,4\}$ so that $r-1 \equiv j \pmod 4$.

We now prove that this solution satisfies the constraints of the dual problem.
First, it is easy to see that $Y^{(t)}$ is Hermitian 
and that $Q_{j}^{(t)} \geq 0$, for any $j \in \{1, \ldots, d\}$.
We prove by induction on $t$ that the rest of the constraints are also satisfied, 
namely all the constraints of the form
\[
Y^{(t)} - \rho_{j}^{(t)} \geq \pt_{\A}(Q_{j}^{(t)}),\quad j \in \{1, \ldots, d\}.
\]
The base case $t = 2$ was considered above.
By the induction hypothesis, and from the fact that $\psi_0 + \psi_1 \geq 0$, 
it holds that
\begin{equation}
(\psi_0 + \psi_1) \otimes Y^{(t)} - 
	(\psi_0 + \psi_1)\otimes\rho_{j}^{(t)}
	\geq (\psi_0 + \psi_1)\otimes\pt_{\A}(Q_{j}^{(t)}).
\end{equation}
From Eq. \eqref{eq:construction}, we have
$\rho_{j}^{(t+1)} = \psi_{0}\otimes\rho_{j}^{(t)}$ if $j \leq 2^{t}$, or 
$\rho_{j}^{(t+1)} = \psi_{1}\otimes\rho_{j-2^{t}}^{(t)}$ if $j > 2^{t}$.
Since $\psi_{0}, \psi_{1} \geq 0$, in either of the two cases we have
\begin{equation}
(\psi_0 + \psi_1) \otimes Y^{(t)} - 
	\rho_{j}^{(t+1)}
	\geq (\psi_0 + \psi_1)\otimes\pt_{\A}(Q_{j}^{(t)}).
\end{equation}
From the set of equations \eqref{eq:transposebells}, 
it is easy to see that
\begin{equation}
\label{eq:pt01}
	\pt_{\A}(\psi_0 + \psi_1) = \psi_0 + \psi_1.
\end{equation}
It follows that
\begin{equation}
(\psi_0 + \psi_1) \otimes Y^{(t)} - 
	\rho_{j}^{(t+1)} 
	\geq \pt_{\A}[(\psi_0 + \psi_1)\otimes(Q_{j}^{(t)})].
\end{equation}
Finally, by the definition of the operators in Eq. \eqref{eq:generalsolution}, we have that
\begin{equation}
	Y^{(t+1)} - \rho_{j}^{(t+1)} 
	\geq \pt_{\A}(Q_{j}^{(t+1)}).
\end{equation} 

In the case where we consider only $k$ of the states we have constructed, 
the value of the program for this solution is equal to
\begin{equation}
	\frac{\tr(Y^{(t)})}{k} = \frac{2^{t-2}\tr(Y^{(2)})}{k} = \frac{7d}{8k}.
\end{equation}
This concludes the proof. \,$\square$

\vspace*{6pt}
It is possible to adapt the construction \eqref{eq:construction} and
\eqref{eq:generalsolution} in order to use a different couple of 
Bell states other than $\psi_0$ and $\psi_1$. 
However, these states are well-suited for a clearer proof, 
due to the Eq. \eqref{eq:pt01}.
\vspace*{6pt}

\begin{cor}
For any $d = 2^{t}$, where $t \geq 4$, there exists a set of $k < d$ 
maximally entangled states in $\complex^{d}\otimes\complex^{d}$ 
that cannot be perfectly distinguished by any LOCC measurement. 
\end{cor}
\vspace*{12pt}
\noindent
{\bf Proof:}
By the above Theorem, when $t \geq 4$, we can construct a set of $k < 2^{t}$ 
states that can be distinguished by any PPT measurement, 
and therefore any LOCC measurement, with only probability 
of success strictly less than $1$. 
In fact, we have that $(7 \cdot 2^{t})/(8 \cdot k) < 1$ 
whenever $t \geq 4$ and $k > (7 \cdot 2^{t})/8$.
\,$\square$

\vspace*{12pt}
Notice that the states generated by the above construction are Bell-diagonal,
like the sets exhibited in \cite{Yu2012} and \cite{Cosentino2013}.
A construction not based on Bell-diagonal states would be needed to generalize 
the result to the case when the dimension is not a power of two.
Unfortunately, the most straightforward generalization, which makes use of the states 
corresponding to the generalized Pauli operators (see \cite{Cosentino2013} for 
a formal definition of these states), leads to weak bounds and does not seem to give
neat analytic solutions of the semidefinite program.

\section{Example}

As an application of our construction, we consider an example where
the two parties are given a state drawn with uniform probability from the following set 
of $k = 15$ orthogonal maximally entangled states in $\complex^{16}\otimes\complex^{16}$:

\begin{align*}
  \rho_1 &= \psi_0 \otimes \psi_0 \otimes \psi_0 \otimes \psi_0,  
    &\rho_2 &= \psi_0 \otimes \psi_0 \otimes \psi_1 \otimes \psi_1, \\
  \rho_3 &= \psi_0 \otimes \psi_0 \otimes \psi_2 \otimes \psi_1,  
    &\rho_4 &= \psi_0 \otimes \psi_0 \otimes \psi_3 \otimes \psi_1, \\         
  \rho_5 &= \psi_0 \otimes \psi_1 \otimes \psi_0 \otimes \psi_0,  
    &\rho_6 &= \psi_0 \otimes \psi_1 \otimes \psi_1 \otimes \psi_1, \\
  \rho_7 &= \psi_0 \otimes \psi_1 \otimes \psi_2 \otimes \psi_1,  
    &\rho_8 &= \psi_0 \otimes \psi_1 \otimes \psi_3 \otimes \psi_1, \\         
  \rho_9 &= \psi_1 \otimes \psi_0 \otimes \psi_0 \otimes \psi_0,  
    &\rho_{10} &= \psi_1 \otimes \psi_0 \otimes \psi_1 \otimes \psi_1, \\
  \rho_{11} &= \psi_1 \otimes \psi_0 \otimes \psi_2 \otimes \psi_1, 
    &\rho_{12} &= \psi_1 \otimes \psi_0 \otimes \psi_3 \otimes \psi_1, \\        
  \rho_{13} &= \psi_1 \otimes \psi_1 \otimes \psi_0 \otimes \psi_0, 
    &\rho_{14} &= \psi_1 \otimes \psi_1 \otimes \psi_1 \otimes \psi_1,\\ 
  \rho_{15} &= \psi_1 \otimes \psi_1 \otimes \psi_2 \otimes \psi_1.
\end{align*}

The probability of distinguishing this set by 
any PPT measurement is less than or equal to $14/15$. 
Examples in higher dimensions can be generated using
the Python script available at \cite{code}.
\bigskip

It is worth noting that the ``Entanglement Discrimination Catalysis'' 
phenomenon, observed in \cite{Yu2012} 
for the set \eqref{eq:example4}, also applies to the set 
of states in the above example and to any set derived 
from our construction.
If Alice and Bob are provided with a maximally entangled state
as a resource, then they are able to distinguish the states
in these sets and, when the protocol ends, 
they are still left with an untouched maximally entangled state.
When $t=2$, the catalyst is used to teleport 
the first qubit from one party to the other, 
say from Alice to Bob. 
Bob can then measure the first two qubits in the standard 
Bell basis and identify which of the four states was prepared. 
Since the third and fourth qubits are not being acted on, 
they can be used in a new round of the protocol.
For the case $t > 2$, let us recall the recursive construction 
of the states $\rho_{j}^{(t)}$ from \eqref{eq:construction}.
Distinguishing between the two cases of the recursion is 
equivalent to distinguishing between two Bell states.
And the base case is exactly the case $t=2$ described above, 
with only one maximally entangled state involved in the catalysis.

\section{Discussion}
In this article we showed an explicit method to generate small sets 
of maximally entangled states that are not perfectly distinguishable by LOCC protocols.
Thus we proved, for the first time, that the dimension of the local subsystems is
not a tight bound on the size of sets of locally indistinguishable orthogonal 
maximally entangled states.

Asymptotically, our construction allows for the cardinality of these sets to be as small 
as $C\cdot d$, where $C$ is a constant less than $1$,
and $d$ is the dimension of each Alice's and Bob's subsystems. 
In particular, we have that $7/8 \leq C < 1$. 
It is possible that this constant can be improved by using 
a different construction or by starting our recursive construction 
from a different base case.
A further improvement would be to show a construction of 
indistinguishable sets with size $o(d)$.
Another open problem is to give a more 
general construction that works even when $d$ is not a power of two.

Finally, the bounds we proved in the paper hold for the class of PPT measurements.
Stronger bounds might hold for the more restricted classes of LOCC or separable measurements.
Navascu\'es showed a hierarchy of semidefinite programs 
for the problem of state distinguishabilty by separable operations \cite{Navascues2008}.
The first level of this hierarchy corresponds to the semidefinite program 
that we studied in this paper. An analysis of higher levels of the hierarchy may lead to stronger 
bounds than the one proved in this article. This idea will be developed in future work.

\section*{Acknowledgements}
We are grateful to John Watrous for many insightful discussions on the problem 
and to Somshubhro Bandyopadhyay for suggesting several improvements to the paper.

This research was supported by Canada's NSERC and the US ARO.


\bibliography{kltd}

\begin{thebibliography}{10}

\bibitem{Bennett99}
Charles~H. Bennett, David~P. DiVincenzo, Christopher~A. Fuchs, Tal Mor, Eric
  Rains, Peter~W. Shor, John~A. Smolin, and William~K. Wootters.
\newblock Quantum nonlocality without entanglement.
\newblock {\em Phys. Rev. A}, 59:1070--1091, 1999.

\bibitem{Walgate00}
Jonathan Walgate, Anthony~J. Short, Lucien Hardy, and Vlatko Vedral.
\newblock Local distinguishability of multipartite orthogonal quantum states.
\newblock {\em Phys. Rev. Letters}, 85:4972, 2000.

\bibitem{Ghosh2001}
Sibasish Ghosh, Guruprasad Kar, Anirban Roy, Aditi Sen(De), and Ujjwal Sen.
\newblock Distinguishability of {B}ell states.
\newblock {\em Phys. Rev. Letters}, 87:277902, 2001.

\bibitem{Walgate2002}
Jonathan Walgate and Lucien Hardy.
\newblock Nonlocality, asymmetry, and distinguishing bipartite states.
\newblock {\em Phys Rev Letters}, 89:147901, 2002.

\bibitem{Horodecki2003}
Michal Horodecki, Aditi Sen(De), Ujjwal Sen, and Karol Horodecki.
\newblock Local indistinguishability: More nonlocality with less entanglement.
\newblock {\em Phys. Rev. Letters}, 90:047902, 2003.

\bibitem{Fan04}
Heng Fan.
\newblock Distinguishability and indistinguishability by local operations and
  classical communication.
\newblock {\em Phys. Rev. Letters}, 92:177905, 2004.

\bibitem{Ghosh04}
Sibasish Ghosh, Guruprasad Kar, Anirban Roy, and Debasis Sarkar.
\newblock Distinguishability of maximally entangled states.
\newblock {\em Phys. Rev. A}, 70:022304, 2004.

\bibitem{Nathanson05}
Michael Nathanson.
\newblock Distinguishing bipartite orthogonal states using {LOCC}: Best and
  worst cases.
\newblock {\em J. Math. Phys.}, 46:062103, 2005.

\bibitem{Watrous2005}
John Watrous.
\newblock Bipartite subspaces having no bases distinguishable by local
  operations and classical communication.
\newblock {\em Phys. Rev. Letters}, 95:080505, 2005.

\bibitem{Owari2006}
Masaki Owari and Masahito Hayashi.
\newblock Local copying and local discrimination as a study for non-locality of
  a set.
\newblock {\em Phys. Rev. A}, 74:032108, 2006.

\bibitem{Yu2011}
Nengkun Yu, Runyao Duan, and Mingsheng Ying.
\newblock Any $2 \otimes n$ subspace is locally distinguishable.
\newblock {\em Phys. Rev. A}, 84:012304, 2011.

\bibitem{Yu2012}
Nengkun Yu, Runyao Duan, and Mingsheng Ying.
\newblock Four locally indistinguishable ququad-ququad orthogonal maximally
  entangled states.
\newblock {\em Phys. Rev. Letters}, 109(2):020506, July 2012.

\bibitem{Yu2012a}
Nengkun Yu, Runyao Duan, and Mingsheng Ying.
\newblock Distinguishability of quantum states by positive operator-valued
  measures with positive partial transpose.
\newblock arXiv:1209.4222, Sep 2012.

\bibitem{Cosentino2013}
Alessandro Cosentino.
\newblock Positive-partial-transpose-indistinguishable states via semidefinite
  programming.
\newblock {\em Phys. Rev. A}, 87:012321, Jan 2013.

\bibitem{Bandyopadhyay2013}
Somshubhro Bandyopadhyay and Michael Nathanson.
\newblock Tight bounds on the distinguishability of quantum states under
  separable measurements.
\newblock {\em Phys. Rev. A}, 88:052313, Nov 2013.

\bibitem{Gottesman2000}
Daniel Gottesman.
\newblock Theory of quantum secret sharing.
\newblock {\em Phys. Rev. A}, 61:042311, Mar 2000.

\bibitem{DiVincenzo2002}
David~P. DiVincenzo, Debbie~W. Leung, and Barbara~M. Terhal.
\newblock Quantum data hiding.
\newblock {\em IEEE Trans. Inform. Theory}, 48(3):580--599, 2002.

\bibitem{Ghosh11}
Somshubhro Bandyopadhyay, Sibasish Ghosh, and Guruprasad Kar.
\newblock {LOCC} distinguishability of unilaterally transformable quantum
  states.
\newblock {\em New J. Phys.}, 13:123013, 2011.

\bibitem{Childs2012}
Andrew~M. Childs, Debbie Leung, Laura Mancinska, and Maris Ozols.
\newblock A framework for bounding nonlocality of state discrimination.
\newblock {\em Comm. Math. Phys.}, 323(3):1121--1153, November 2013.

\bibitem{Duan2009}
Runyao Duan, Yuan Feng, Yu~Xin, and Mingsheng Ying.
\newblock Distinguishability of quantum states by separable operations.
\newblock {\em IEEE Trans. Inform. Theory}, 55(3):1320--1330, 2009.

\bibitem{Gurvits2003}
Leonid Gurvits.
\newblock Classical deterministic complexity of {E}dmonds' problem and quantum
  entanglement.
\newblock In {\em Proceedings of the thirty-fifth annual ACM Symposium on
  Theory of Computing}, STOC '03, pages 10--19, New York, NY, USA, 2003. ACM.

\bibitem{Gharibian2008}
Sevag Gharibian.
\newblock Strong {NP}-hardness of the quantum separability problem.
\newblock {\em Quantum Information and Computation}, 10:No.3\&4, 343--360,
  2010.

\bibitem{Rains00}
Eric Rains.
\newblock A semidefinite program for distillable entanglement.
\newblock {\em IEEE Trans. Inform. Theory}, 47(7):2921--2933, 2001.

\bibitem{code}
Alessandro Cosentino and Vincent Russo.
\newblock Code repository.
\newblock \url{https://bitbucket.org/acosenti/ppt-sdp-paper}.

\bibitem{Navascues2008}
Miguel Navascu\'es.
\newblock Pure state estimation and the characterization of entanglement.
\newblock {\em Phys. Rev. Lett.}, 100:070503, Feb 2008.

\end{thebibliography}
\bibliographystyle{unsrt}

\end{document}